\documentclass[aps,nofootinbib,superscriptaddress,twocolumn,preprintnumbers,groupedaddress]{revtex4}
 \usepackage{amstext,amsmath,amssymb,amsfonts,bbm}
\usepackage[latin1]{inputenc}
\usepackage{fancyhdr}
\usepackage{graphicx}
\usepackage{hyperref}
\usepackage{epstopdf}
\usepackage{amsthm}

\def\be{\begin{equation}}
\def\ee{\end{equation}}
\def\bes{\begin{eqnarray}}
\def\ees{\end{eqnarray}}

\def\f{\frac}

\begin{document}

\title{\large \bf Thermal time and Tolman-Ehrenfest effect:\\ ``temperature as the speed of time"}

\author{Carlo Rovelli, Matteo Smerlak}
\affiliation{Centre de Physique Th\'eorique de Luminy\footnote{Unit\'e mixte de recherche (UMR 6207) du CNRS et des Universit\'es de Provence (Aix-Marseille I), de la M\'editerran\'ee (Aix-Marseille II) et du Sud (Toulon-Var); laboratoire affili\'e \`a la FRUMAM (FR 2291).}, Case 907, F-13288 Marseille, EU}

\date{\small\today}

\begin{abstract}\noindent
The notion of \emph{thermal time} has been introduced as a possible basis for a fully general-relativistic thermodynamics. Here we study this notion in the restricted context of stationary spacetimes. We show that the Tolman-Ehrenfest effect (in a stationary gravitational field, temperature is \emph{not} constant in space at thermal equilibrium) can be derived very simply by applying the equivalence principle to a key property of thermal time: at equilibrium, temperature is the rate of thermal time with respect to proper time -- the `speed of (thermal) time'. Unlike other published derivations of the Tolman-Ehrenfest relation, this one is free from any further dynamical assumption, thereby illustrating the physical import of the notion of thermal time.

\end{abstract}

\maketitle

\section{Introduction}

The notion of \emph{thermal time} was put forward by Connes and one of the authors \cite{Rovelli:1993ys,Connes:1994hv} as a basis for a fully general-relativistic thermodynamics -- a problem which is still open \cite{Rovelli:1993ys,Smolin:1982jt,Montesinos:2000zi,Rovelli:2009ks}. In a nutshell, the thermal time of a system is the natural flow induced \emph{by its statistical state} on its algebra of observables. Unlike other notions of time, which are tied to the existence of a spacetime metric, thermal time remains meaningful in the quantum gravitational regime \cite{Rovelli}. 

So far, however, thermal time has remained a rather abstract notion, with few concrete applications \cite{Rovelli:1993ysBIS,Martinetti:2002sz,Martinetti:2004ji,Tian:2005yj,Martinetti:2008ja}, and the doubt is legitimate whether it has much physical content. The purpose of this note is to show that it does, and actually provides the most economical means to describe the influence of gravity on temperature. We do so by deriving the Tolman-Ehrenfest law from a straightforward application of the equivalence principle to the thermal time of non-relativistic equilbrium states.

Recall that the Tolman-Ehrenfest effect is the fact that temperature is {\it not} constant in space at equilibrium,  in the presence of gravity \cite{Tolman:1930zz,TolmanEhrenfest}. In a stationary spacetime with timelike Killing vector field $\xi$, the temperature  $T$ satisfies instead the Tolman-Ehrenfest relation
\be
T\Vert\xi\Vert=\textrm{const},
\label{goo}
\ee
where
 $\Vert\xi\Vert=\sqrt{g_{ab}\xi^a\xi^b}$ is the spacetime norm of $\xi$. In the Newtonian limit, this corresponds to a temperature gradient 
\be 
\f{\nabla T}{T}=\f{\vec{g}}{c^2}, \label{te}
\ee 
where $\vec g$ is the Galilean acceleration of gravity. In few simple words, a vertical column of fluid at equilibrium is hotter at the bottom. (Of course, this is a tiny $1/c^2$ relativistic effect, negligible in most practical situations;\footnote{On the surface of the Earth, $\f{\nabla T}{T}=10^{-18} cm^{-1}$.} from a theoretical persepective, nonetheless, it is very significant, for it is constitutes a bridge between thermodynamics and general relativity.)

Tolman and Ehrenfest's original derivation of this relation was based on a number of dynamical assumptions, including the validity of the Einstein equations \cite{Tolman:1930zz,TolmanEhrenfest}. Other authors later simplified their derivation by stripping these down to what appeared their bare essential: the relativistic equivalence between mass and energy \cite{Landau,Balazs,Ehlers,Tauber,Balazs2,Buchdahl,Ebert,Stachel}. (We recall such a simple derivation in Appendix A.) But the strength of thermodynamics as we usually know it is not to rely on few dynamical assumptions: it is to rely on \emph{no} dynamical assumption.

Here, we show that the Tolman-Ehrenfest law follows from applying the equivalence principle to thermal time, and nothing else. This entails in particular that the geometric properties of the thermal time flow of an equilibrium state remain valid in a curved spacetime. The first of these properties states that the thermal time flow generates a Killing symmetry of the metric. The second is that the ratio between the flow of thermal time and the flow of proper time -- the `speed of (thermal) time' -- is the temperature. In a stationary spacetime, the relation \eqref{goo} follows immediately from these two condtions.

Besides demonstrating the truly thermodynamical nature of the Tolman-Ehrenfest effect, this result illustrates in our opinion the import and effectiveness of the notion of thermal time -- its physical content.

\section{Thermal time}
We start by recalling the mathematical definition of thermal time, in the simplified setting of classical Hamiltonian mechanics. (The full quantum version is recalled for completeness in Appendix B.)  A general relativistic {\it statistical} system can be described by a Poisson algebra $\cal A$ of observables $A$ on a phase space ${\cal S}$. Statistical states $\rho$ are normalized\footnote{In the sense that $\int_{{\cal S}}ds\ \rho(s)=1$.} positive functions on ${\cal S}$, interpreted as probability densities on ${\cal S}$. 

Given a statistical state $\rho$, we define the \emph{thermal time flow} $\alpha^\rho_\tau:\cal A\rightarrow\cal A$ as the Poisson flow of $(-\ln\rho)$ in $\cal A$. That is 
\be 
          \frac{d\alpha^\rho_\tau(A)}{d\tau}=- \{A, \ln \rho  \}.
\ee
where the r.h.s.~is the Poisson bracket. 

The import of this definition is that it relies only on the statistical state and on the Poisson structure of the system, and makes no reference to a kinematical time variable.  For this reason, it continues to make sense in a fully general-relativistic context, where preferred notions of time may be absent.

Consider now a non-relativistic Boltzmann-Gibbs \emph{equilibrium} state  $\rho_T$, describing thermal equilibrium at temperature $T$ in the canonical ensemble,
\be
          \rho_T= Z^{-1} e^{-\frac H{kT}},
          \label{bg}
\ee
where $H$ is the energy, $k$ the Botzmann constant and $Z=\int_{{\cal S}}ds\ e^{-\frac H{kT}}$. The thermal time flow of $\rho_T$ is 
\be
\frac{d\alpha^{\rho_T}_\tau(A)}{d\tau}=- \{A, \ln \rho_T  \}=\frac1{kT}\ \{A, H\}
=\frac1{kT}\ \frac{dA}{dt}
\label{flow}
\ee
Hence the thermal time $\tau$ of an equilibrium state at temperature $T$ and the Newtonian mechanical time $t$ are related by 
\be
             \f{d}{d\tau} = \frac1{kT}\ \f{d}{dt}.
             \label{ttt}
\ee
In other words, the thermal time of a non-relativistic equilibrium state satisfies the following two properties: 
\begin{enumerate}
\item[\em (i)] The flow $\f{d}{d\tau}$ has a geometric action on spacetime: it acts on observables in the same manner as a vector field $\xi=\frac1{kT}\ \f{d}{dt}$. Furthermore $\xi$ is a Killing symmetry of the (Galilean) spacetime: it generates a one-parameter group of timelike isometries. 
\item[\em (ii)]  The ratio between the flow of thermal time $\tau$ and the flow of mechanical time $t$ is the temperature $kT$. 
\end{enumerate}

Property $(i)$ states that, \emph{for an equilibrium state}, the thermal time flow is proportional to the kinematical time flow. Property $(ii)$ provides a new and remarkable definition of temperature: temperature is the   ``speed" of thermal time, namely the ratio between the flow of thermal time and the flow of mechanical time.   

A crucial fact pointed out in \cite{Martinetti:2002sz, Martinetti:2008ja} is that {\em (ii)} can be interpreted locally: at any given spacetime point, temperature is given by the ratio between the two flows at that point.  In the next section, we use this observation to derive the Tolman-Ehrenfest law. 

\section{Stationary spacetimes}

Consider now more generally a macroscopic system, say a gas, in a \emph{stationary} spacetime.  The phase space $\cal S$ of such a system can be thought of covariantly as the set of solutions  $\vec{x}_n(t)$ of the equations of motion of all the particules of the gas. Observables are functions of these trajectories. Among them are the {\it local} observables $A_x$, which depend on the positions and momenta of the particles in a neighborhood of the spacetime point $x=(\vec x, t)$ (for instance the density). 


Let $\rho$ be an \emph{equilibrium} state and $\alpha_\tau^\rho$ its thermal time flow.  We now make the physical hypothesis that the equivalence principle can be applied to thermal time and the two properties {\em (i,ii)}. The first then becomes 
\begin{enumerate}
\item[\em (i')] The thermal time flow $\alpha^{\rho}_{\tau}$ has a geometric action on local observables $A_x$ according to
\be
\frac{d\alpha^{\rho}_\tau(A)}{d\tau}=\mathcal{L}_{\xi^\rho}A_{x},
\label{tk}
\ee
where the Lie derivative $\mathcal{L}_{\xi^\rho}$ acts on $A_x$ seen as a function on spacetime, and the vector field $\xi^\rho$ generates a timelike symmetry of spacetime, that is, is a timelike Killing field for the stationary metric.\end{enumerate}
The second, which gives the temperature, now reads 
\begin{enumerate}
\item[\em (ii')] At every space point, that is along every stationary timelike curve, temperature is the ratio of the thermal time flow to proper time $s$,
\be
\xi^\rho=\f{1}{kT}\f{d}{ds}.
\label{defrel}
\ee
\end{enumerate}

Now, observe that at each point a timelike Killing field is tangent to $d/ds$ along a stationary timelike curve, but in general the ratio of their norms is \emph{not} constant in space.  By taking the norm of the last equation, we have indeed 
\be
\Vert\xi^\rho\Vert=\f{1}{kT}. 
\ee
Replacing $\xi^\rho$ by any other tangent Killing vector (which must be globally proportional to $\xi^\rho$), we get the Tolman-Ehrenfest law \eqref{goo}. Thus, the influence of gravity on thermal equilibrium can be obtained directly by applying the equivalence principle to the two properties  {\em (i)} and {\em (ii)} of thermal time.

\section{Discussion}

1. Thermal equilibrium states are the final states of irreversible processes. In standard statistical mechanics, such states can be characterized in a number of ways: stochastically, by the condition of {\it detailed balance} of microscopic probability fluxes; dynamically, by a condition of {\it stability} under small perturbations of the Hamiltonian; thermodynamically, by a condition of {\it passivity}\footnote{Passivity refers to the impossibility to extract work from cyclic processes in a system at thermal equilibrium -- Kelvin's formulation of the second law.}; information-theoretically, by the {\it maximization of entropy}; quantum mechanically, by the periodicity of correlation functions in imaginary time, aka the KMS condition; and so on.  When moving to curved spacetimes, these characterizations tend to become problematic, because of the effect of gravity. In the case of {\it stationary} spacetimes, several generalizations of the condition of thermal equilibrium have been studied \cite{Ebert,Landau, Balazs, Ehlers,Tauber,Balazs2,Buchdahl}. Conditions {\em (i',ii')} can be taken as a general characterization of thermal equilibrium in a stationary spacetime. They reduce to the standard Boltzmann-Gibbs ansatz in the non-relativistic context, but remain valid on general stationary spacetimes. 

2. Notice that condition $(i)$ consists of two parts. First, it states that the thermal time induced by a thermal state $\rho$ is geometric, that is, there exist a vector field $\xi^\rho$ for which \eqref{tk} holds. This is a highly non-trivial condition on the state, and understanding the underlying mathematics is an interesting open problem \cite{Borchers:2000fk,geometricmodular}. (In the context of algebraic quantum field theory, it has even been proposed that this \emph{condition of geometric modular action} can be used as a ``criterion for selection physically interesting states on general spacetimes"
\cite{geometricmodular}.)  Second, condition {\em (i)} states that $\xi^\rho$ is Killing. This means that the state shares the same time-translation symmetry as the underlying space-time, giving overall stationarity.  Having the thermal time flow proportional to a timelike Killing field is a physical property which we propose captures the condition of equilibrium. 

3. In this note we have studied the notion of thermal time in the context of stationary spacetimes.
The motivation for the introduction of this notion, though, was to provide a notion of time flow in the physical situation where the gravitational field itself is thermalized (or is in a quantum state) and therefore the thermal (or quantum) superposition of geometries makes conventional notions of time meaningless. For statistical states at, or approaching, equilibrium on a fixed geometry, the two notions match and this matching is part of the condition of equilibrium itself, since the gravitational field must be stationary with respect to the thermal time in order for the entire system to be stationary.
But the notion of thermal time retains its validity also in the context where a classical background metric is not available. The ``thermal time hypothesis" of \cite{Rovelli:1993ys,Connes:1994hv}, indeed, postulates that {\it the thermal time governing the thermodynamics of a macroscopic system described by a given statistical state is this flow}. This might allow to define thermodynamics also in the context where space-time is not defined, because of thermal or quantum fluctuations of the geometry. We do not address this issue here.

\section{Conclusion}

Summarizing, we have shown that the non-relativistic properties of thermal time can be directly generalized to stationary spacetimes, leading immediately to the Tolman-Ehrenfest effect. The key feature of thermal time used in this derivation is the fact that the ratio between the the flow of thermal time and the flow of proper time is the temperature. This fact appears somewhat tautological in flat space, as it amounts to fixing an arbitrary unobservable scale for the thermal time.  But it becomes consequential in a curved spacetime, where the norm of the Killing field, and hence the thermal time flow, varies from point to point. While the global scale of thermal time remains arbitrary, the ratio between its flow in different spacetime points (with respect to proper time) is physically meaningful. 

Thus, the Tolman-Ehrenfest law can be stated very simply as: the stronger the gravitational potential, the faster the thermal time flow with respect to proper time, and hence the higher the temperature. It is amusing in this respect to notice that the expression ``thermal time" is also used in biology to indicate the linear relationship between development rate and temperature which is widely observed among plants and ectotherms \cite{Bonhomme}. For them too -- the hotter, the faster.

\appendix
\section{A simple derivation of the Tolman-Ehrenfest effect}

Several intuitive arguments can be used to make physical sense of the Tolman-Ehrenfest effect. Here is one, which makes use of $E=mc^2$ and the equivalence of inertial and gravitational mass. Equilibrium between two systems happens when the total entropy is maximized \be
dS=dS_1+dS_2=0.\ee 
If a heat quantity $dE_1$ leaves the first system, and the same quantity $dE_2=dE_1$ enters the second system, then 
\be
dS_1/dE_1-dS_2/dE_2=0.
\ee
 Since $T:=dS/dE$, we obtain $T_2=T_1$. 
 
 However, if the two systems are at different gravitational potentials (in the Newtonian limit), the amount of energy $dE_1$ leaving, say, the upper one, is not the amount of energy $dE_2$ entering the lower one. Indeed, $E=mc^2$ and the equality of inertial and gravitational mass imply that any form of energy has a gravitational mass, and ``falls". Hence $dE_2$ is $dE_1$ increased by the potential energy $m \Delta\Phi$, where $\Phi$ is the gravitational potential:
 \be
 dE_2=dE_1(1+\Delta \Phi/c^2),
 \ee
 which yields immediately (\ref{te}).

\section{Thermal time hypothesis for general relativistic quantum systems}

A general covariant quantum system can be described by an algebra $\cal A$ of observables $A$ (a von Neumann algebra) and a states on $\cal A$.  For instance, in quantum gravity the pure states can be given by the solutions $\psi$ of the Wheeler-DeWitt equation, and observables by self-adjoint operators on a Hilbert space defined by these solutions (see for instance  \cite{Rovelli}). \emph{Statistical} states are incoherent superpositions of such states, and can be defined as positive linear functionals on $\cal A$. 

Given a (faithful, normal) state $\rho$, the \emph{thermal time flow} $\alpha_\tau^\rho$ is the Tomita flow of the state $\rho$ in $\cal A$ \cite{Connes:1994hv}. It satisfies in particular 
\be
\rho(\alpha_\tau^\rho A)=\rho(A). 
\label{rho}
\ee

This flow depends on the state, but the flows generated by different states are equivalent up to inner automorphisms in $\cal A$  \cite{al,albook}. Therefore we can further define an \emph{outer thermal time flow} $\beta_\tau$, as the flow  $\alpha^\rho_\tau$ up to inner automorphisms. Remarkably, this is state independent.



\providecommand{\href}[2]{#2}\begingroup\raggedright\endgroup


\end{document}